\documentclass[lettersize,journal]{IEEEtran}
\usepackage{amsmath,amsfonts}
\usepackage{algorithmic}
\usepackage[ruled, noend]{algorithm2e}
\usepackage{array}
\usepackage{textcomp}
\usepackage{stfloats}
\usepackage{url}
\usepackage{verbatim}
\usepackage{graphicx}
\usepackage{cite}
\usepackage{bm}
\usepackage{subfigure}

\usepackage{clipboard}
\usepackage{enumerate}
\usepackage[normalem]{ulem}
\usepackage{enumitem}

\newcommand{\sTx}{\mathrm{Tx}}
\newcommand{\sRx}{\mathrm{Rx}}
\newcommand{\ssc}{\mathrm{sc}}

\newcommand{\sGR}{\mathrm{GR}}
\newcommand{\sBFI}{\mathrm{BFI}}
\newcommand{\sSel}{\mathrm{sel}}

\newcommand{\srsvd}{\mathrm{rsvd}}

\newcommand{\sG}{\mathrm{G}}

\newcommand{\iu}{\mathrm{i}}
\newcommand{\dd}{\mathrm{d}}
\newcommand{\Diag}{\mathbf{Diag}}

\newcommand{\beq}{\begin{equation}}
\newcommand{\eeq}{\end{equation}}

\usepackage{color}

\newtheorem{proposition}{\bf Proposition}

\renewcommand{\mod}{\mathrm{mod}}

\ifodd 0
\newcommand{\rev}{\textcolor[rgb]{0,0.,0.7}}
\else
\newcommand{\rev}{\textcolor[rgb]{0,0.,0.}}
\fi

\ifodd 1
\newcommand{\addrev}{}
\newcommand{\minirev}{}

\newcommand{\huv}{}

\else
\newcommand{\addrev}{}

\newcommand{\huv}{}

\fi

\begin{document}

\title{Efficient Beamforming Feedback Information-Based Wi-Fi Sensing by Feature Selection}

\author{
\IEEEauthorblockN{
\normalsize{Xin~Li},~\IEEEmembership{\normalsize Member,~IEEE},
\normalsize{Jingzhi~Hu},~\IEEEmembership{\normalsize Member,~IEEE},
and~\normalsize{Jun~Luo},~\IEEEmembership{\normalsize Fellow,~IEEE}
\vspace{-2.em}}
\thanks{© 2024 IEEE.  Personal use of this material is permitted.  Permission from IEEE must be obtained for all other uses, in any current or future media, including reprinting/republishing this material for advertising or promotional purposes, creating new collective works, for resale or redistribution to servers or lists, or reuse of any copyrighted component of this work in other works.

This research is supported in part by National Research Foundation (NRF) Future Communications Research \& Development Programme (FCP) grant FCP-NTU-RG-2022-015.

X. Li, J. Hu, and J. Luo are with the College of Computing and Data Science, Nanyang Technological University, Singapore.~(email: \{l.xin,~jingzhi.hu,~junluo\}@ntu.edu.sg)}
}



\maketitle
\begin{abstract}
Wi-Fi sensing leveraging plain-text beamforming feedback information~(BFI) in multiple-input-multiple-output~(MIMO) systems attracts increasing attention.
However, due to the implicit relationship between BFI and the channel state information~(CSI), quantifying the sensing capability of BFI poses a challenge in building efficient BFI-based sensing algorithms.
In this letter, we first derive a mathematical model of BFI, characterizing its relationship with CSI explicitly, and then develop a closed-form expression of BFI for $2\times2$ MIMO systems.
To enhance the efficiency of BFI-based sensing by selecting only the most informative features, we quantify the sensing capacity of BFI using the Cramer-Rao bound~(CRB) and then propose an efficient CRB-based BFI feature selection algorithm.
Simulation results verify that BFI and CSI exhibit comparable sensing capabilities and that the proposed algorithm halves the number of features, \rev{reducing 20\% more parameters than baseline methods, at the cost of only slightly increasing positioning errors.}

\end{abstract}


\section{Introduction}
\label{sec:intro}

Recently, ubiquitous sensing has been widespread
to enhance user experiences, providing real-time information on positions, gestures, activities, etc.
\Copy{R3-1-6}{As one of the most promising techniques \huv{for ubiquitous sensing, Wi-Fi sensing leverges the channel state information~(CSI), which is readily provided by existing Wi-Fi infrastructure}. 
However, \huv{current} CSI-based Wi-Fi sensing requires special modification\huv{s of} Wi-Fi network interface cards~(NICs) to extract CSI at the physical layer, posing challenges for its broad implementation \huv{in practice}~\cite{hu2023muse}.

Fortunately, the \emph{explicit beamforming mechanism} used in prevalent Wi-Fi 802.11ac/ax networks for multiple-input-multiple-output~(MIMO) communications provides an accessible alternative for Wi-Fi sensing~\cite{hu2023password}}.
\huv{In particular}, a user device~(UD) feedbacks a compressed form of downlink CSI, termed \emph{beamforming feedback information}~(BFI) in Wi-Fi standards, to the access point~(AP), helping the AP conduct effective downlink beamforming to it.
As BFI contains channel information between the AP and UD and is transmitted in easily-accessible plain text packets, it offers a straightforward solution for ubiquitous Wi-Fi sensing.

Existing works on BFI-based sensing have demonstrated its practicability~\cite{Itahara2022Acc_Beamforming, Wu2023SIGCOMM_Enabling,hu2023muse} and the existence of \huv{positional} information in BFI, including angle of arrival~(AoA), angle of departure~(AoD), and distance~\cite{Jiang2022WCL_On}.
The sensing algorithms developed in existing works are based on the complete BFI or the inverse transformation of BFI to CSI, and their efficiency is hampered by the large size of BFI \huv{due to the large number of subcarriers used} in prevalent Wi-Fi networks.
To enhance the efficiency of BFI-based sensing and facilitate its implementation in edge devices with strict resource limitation\huv{s}, selecting only the most informative BFI features is crucial.
\Copy{R3-1-7}{However, the intricate relationship between CSI and BFI poses a challenge in evaluating the sensing capability of BFI \huv{and designing efficient feature selection algorithms}}.

To handle \huv{this challenge}, we derive a mathematical model of BFI, \huv{characterize} its relationship with CSI, and develop a closed-form BFI expression for $2\times 2$ MIMO systems.
Based on the BFI model, we analyze the Cramer-Rao bound~(CRB) of BFI and propose a \huv{Gaussian-kernel-based} approximation method to \huv{measure} the sensing capability of BFI \huv{features}.
Based on the CRB, we propose an efficient BFI feature selection algorithm to reduce the number of BFI features needed \huv{for high-precision} positioning.
Simulation results show that BFI and CSI exhibit comparable sensing capabilities and that the proposed algorithm can halve the number of BFI features \huv{at the cost of} a slight increase in positioning errors.

The rest of this paper is organized as follows: Sec.~\ref{sec:preliminaries} derives the model of BFI, and Sec.~\ref{sec: aprox BCRB} \huv{proposes} the Gaussian-kernel-based CRB approximation and the efficient BFI feature selection algorithm. 
Simulation evaluations are provided in Sec.~\ref{sec:simu and evalu}, and a conclusion is drawn in Sec.~\ref{sec:Conclusion}.

\vspace{-.5ex}
\section{Derivation of BFI Model} \label{sec:preliminaries}

We first describe the basic of CSI in MIMO systems, focusing on Wi-Fi sensing scenarios. 
Subsequently, we establish the BFI model based on its transformation from CSI, and derive the closed-form BFI expression in $2\times 2$ MIMO systems.
\vspace{-.5ex}
\subsection{CSI in Wi-Fi Sensing Systems}\label{ssec:csi-wifi}
In this section, we model 
an \addrev{$N\times M$} MIMO Wi-Fi system with $N_{\ssc}$ subcarriers, where the UD has $N$ Rx antennas, and the AP has \addrev{$M$ Tx antennas}.
Our primary focus is to illuminate the relationship between BFI and CSI.

Based on~\cite{Jiang2022WCL_On}, upon receiving a Wi-Fi frame, CSI data can be obtained.
Specifically, for the $k$-th subcarrier~($k\in\{1,...,N_{\ssc}\}$), the CSI is represented by a matrix \addrev{$\bm H_{k}\in \mathbb C^{N\times M}$}, whose $(n, m)$-th element can be expressed as
\begin{align}
\label{equ: CSI channel}
[\bm H_{k}]_{n,m} = &\sum_{l=1}^{L} \alpha_l e^{-\frac{\iu 2\pi}{\lambda_k} \cdot ( \sin(\varphi_l) \cdot (n-1)\cdot \Delta d_{\sRx} + \sin(\rho_l) \cdot (m-1)\cdot \Delta d_{\sTx})} \nonumber \\
& \cdot e^{-\iu 2 \pi d_l f_k/c} + \epsilon,
\end{align}
\Copy{R3-1-1}{where $c$ is the speed of light,
$n$ and $m$ indicate the indices of Rx and Tx antennas, 
$k$ is the subcarrier index, 
$l$ denotes the path index,
$L$ is the path number,
$f_k$ and $\lambda_k$ are the frequency and wavelength of subcarrier $k$,
and $\Delta d_{\sTx}$ and $\Delta d_{\sRx}$ are the antenna spacings,
$\epsilon\sim\mathcal{CN}(0,\delta^2)$ is a random gain accounting for the thermal noise, interference, and environmental dynamics.
Besides, $\alpha_l$, $d_l$, $\varphi_l$ and $\rho_l$ are the amplitude attenuation, distance, AoA, and AoD for path $l$, respectively}.

\begin{figure*}[b]
\vspace{-.5em}
\centering
\rule{\linewidth}{0.4pt}
\begin{equation}
\label{equ: D and G}
\small
\bm D_i=\left[\begin{array}{ccccc}
\bm I_{i-1} & 0 & 0 & \cdots & 0 \\
0 & e^{\iu \phi_{i, i}} & 0 & \cdots & 0 \\
0 & 0 & \cdots & 0 & 0 \\
0 & 0 & 0 & e^{\iu\phi_{\addrev{M-1}, i}} & 0 \\
0 & 0 & 0 & 0 & 1
\end{array}\right],~
\bm G_{\ell, i}\!(\psi_{\ell, i})=\!\left[\begin{array}{ccccc}
\bm I_{\ell-1} & 0 & 0 & \cdots & 0 \\
0 & \cos \psi_{\ell,i} & 0 & \sin \psi_{\ell, i} & 0 \\
0 & 0 & \bm I_{i-\ell-1} & 0 & 0 \\
0 & -\sin \psi_{\ell, i} & 0 & \cos \psi_{\ell, i} & 0 \\
0 & 0 & 0 & \cdots & \bm I_{\addrev{M-i}}
\end{array}\right]
\end{equation}
\end{figure*}

\subsection{Transformation from CSI to BFI} \label{ssec: trans CSI to BFI}

\Copy{R3-2}{Based on IEEE 802.11n/ac/ax standard~\cite{IEEE80211acD31}, for each subcarrier, we omit the subcarrier index $k$ and denote the BFI after quantization by $\check{\bm \theta}$ and the CSI by $\bm H\in \mathbb C^{N\times \addrev{M}}$.
The mapping from $\bm H$ to $\check{\bm \theta}$ involves four steps: \addrev{i) $\bm V = \bm F_{\srsvd}(\bm H)$; ii) $\tilde{\bm V} = \bm R(\hat{\bm V})$; iii) ${\bm \theta} = \bm g_{\sGR}(\tilde{\bm V})$, and iv) $\check{\bm \theta} = \bm q({\bm \theta}; b)$.}

The $\bm F_{\srsvd}(\cdot)$ obtains the right-singular matrix of SVD,
and the SVD of $\bm H$ can be expressed as $\bm H =  \bm U\bm \varGamma \bm V^{*}$.
Then, the right-singular matrix $\bm V$, referred to as the \emph{steering matrix}, is
\beq
\bm V =  \bm F_{\srsvd}(\bm H)=  \addrev{\bm H^*\bm U(\bm\varGamma^{\dagger})^*},
\eeq
where \addrev{$\bm U$, $\bm V$, and $\bm \varGamma$ are the unitary left- and right-singular matrix and the diagonal matrix of singular values of $\bm H$, respectively. Superscript $\dagger$ denotes the pseudo-inverse.}

\addrev{Truncation (if $N\leq M-1$) or zero-padding (if $N>M$) is applied to process $\bm V$ along the column dimension, resulting in $\hat{\bm V}\in\mathbb C^{M\times N}$.}
\Copy{R3-1-2}{Then, function $\bm R(\hat{\bm V})$ rotates each column of $\hat{\bm V}$ so that the output matrix has real-valued last row}, i.e.,
\beq
\label{eq: v_hat}
\tilde{\bm V} = \bm R(\hat{\bm V}) = \hat{\bm V} \Diag(e^{-\iu\angle[\hat{\bm V}]_{\addrev{M},1}},...,e^{-\iu\angle[\hat{\bm V}]_{\addrev{M, N}}}),
\eeq 
which is a prerequisite for $\tilde{\bm V}$ to be handled by $\bm g_{\sGR}(\cdot)$. 
Specifically, $\bm g_{\sGR}(\tilde{\bm V})$ returns an angular parameter vector $\bm\theta = (\bm \phi, \bm \psi)$ for a sequence of Givens rotation matrices that can decompose $\tilde{\bm V}\in\mathbb C^{\addrev{M}\times N}$.
Though $\bm g_{\sGR}(\tilde{\bm V})$ is an iterative transformation~\cite{Kim2006VTC_Efficient}
and cannot be expressed as an explicit closed-form, it can be defined by its inverse function:
\beq
\tilde{\bm V} = \bm g^{-1}_{\sGR}(\bm \theta) = \!\prod_{i=1}^{\addrev{\min(N,M-1)}} \! \Big(
\bm D_i\! \prod_{\ell = i+1}^{\addrev{M}} \! \bm G^{\top}_{\ell,i}
\Big)\bm I_{\addrev{M\times N}},
\label{eq: v_hat_recover}
\eeq
where $\bm D_i$ and $\bm G_{\ell, i}$ are defined in~\eqref{equ: D and G}.
We can observe that $\bm \psi$ and $\bm \phi$ contain distinct information related to steering matrix $\bm V$:  $\bm \psi$ determines the Givens rotation matrices that rotates $\tilde{\bm V}$ into diagonal matrices, while $\bm \phi$ records the phase offsets for the rotations.
\Copy{R2-3-3}{\addrev{Besides, $\bm\phi\!\in\![0,2\pi]^{N_{\sBFI}/2}$, $\bm\psi\!\in\![0,\pi/2]^{N_{\sBFI}/2}$, and thus $\bm\theta$ comprises $N_{\sBFI}=2MS-S^2-S$ elements with $S=\min(N,M-1)$.}}
\Copy{R1-2nd-Q2}{\addrev{As derived in Proposition~\ref{prop: 1}, the number of independent real variables of $\tilde{\bm V}$ is equal to $N_{\sBFI}$, thus $\bm \theta$ contains all the information of \huv{$\hat{\bm V}$ and} $\tilde{\bm V}$ and the compression by $\bm g_{\sGR}(\cdot)$ does not loss any information.}}
\huv{We refer to elements of $\bm \theta$ as \emph{BFI elements}, or \emph{BFI features} in the context of sensing.}

\begin{proposition}
\label{prop: 1}
\addrev{Assume $\bm H\in\mathbb C^{N\times \addrev{M}}$, and $\hat{\bm V}\in\mathbb C^{\addrev{M}\times N}$ being the resized right-singular matrix of $\bm H$. The number of independent variables of $\hat{\bm V}$ and $\tilde{\bm V}$ are $2MS-S^2-S$}.
\end{proposition}

\begin{IEEEproof}
\addrev{If $N\leq M-1$, $\hat{\bm V}$ contains $2MN$ real-valued elements and satisfies $\hat{\bm V}^*\hat{\bm V}=\bm I_{N\times N}$ due to $\bm V$ being a unitary matrix. We can obtain $N^2+N$ independent constraints in the real domain from the upper triangular matrix. Therefore, the number of independent real-valued elements in $\hat{\bm V}$ is $2MN-N^2-N$. If $N\geq M$, $\hat{\bm V}$ contains $2M^2$ real-valued elements and satisfies $\hat{\bm V}^*\hat{\bm V}=\bm I_{M\times M}$, leading to $M^2+M$ independent constraints. Therefore, the number of independent real-valued elements in $\hat{\bm V}$ is $M^2-M$.
With $S=\min(N,M-1)$ and $\bm R(\hat{\bm V})$ not changing the number of independent variables, the number in $\hat{\bm V}$ and $\tilde{\bm V}$ under both cases can be expressed as $2MS-S^2-S$.}
\end{IEEEproof}}

Finally, $\bm q(\cdot)$ denotes the quantization function with $b$ indicate the resolution.
Since $b$ can be adjusted to ensure enough precision, we focus on BFI $\bm \theta$ before quantization in this paper.

\vspace{-2ex}
\subsection{Closed-Form BFI Expression for $2\times 2$ MIMO Systems} \label{ssec:close form BFI of 2-2}
For a general system, the relationship between precise BFI $\bm \theta$ and CSI $\bm H$ cannot be expressed in a closed-form due to the inherent complexity of the SVD and the Givens rotation. 
Nevertheless, for the special case $M=N=2$, we can derive the closed-form relationship between $\bm \theta$ and $\bm H$, which is of practical value due to the wide application of $2\times2$ MIMO systems.
Thus, without loss of generality, the CSI matrix for an arbitrary subcarrier can be expressed in polar form as
\beq
\label{eq: 2-2 CSI}
\bm H=\left[\begin{array}{cc}
\alpha_{11} \cdot e^{-\iu 2 \pi t_{11}} & \alpha_{12} \cdot e^{-\iu 2 \pi t_{12}} \\
\alpha_{21} \cdot e^{-\iu 2 \pi t_{21}} & \alpha_{22} \cdot e^{-\iu 2 \pi t_{22}} \\
\end{array}\right].
\eeq

Denoting the right-singular matrix and the diagonal eigenvalue matrix of $\bm H$ as $\bm V$ and $\bm \varGamma$, respectively:
\beq
\label{equ: general equ}
\bm H^*\bm H = \bm V\bm\varGamma\bm\varGamma\bm V^*,
\eeq
which comprises a set of $4$ equality constraints accounting for the $2\times 2$ elements on both side.
Since $\bm V$ is unitary matrix and $\bm \varGamma$ is a real-valued diagonal matrix, they contain $4$ independent real-valued variables in total.
\Copy{R3-3-1}{\addrev{Therefore, we can solve $\bm V$ explicitly using the four equality constraints in~\eqref{equ: general equ}.}
\addrev{Moreover, based on~\cite{Kim2006VTC_Efficient}, matrix $\tilde{\bm V}$, the resulting matrix of an $2\times 2$ unitary matrix after rotating its last row to real numbers, can be generally expressed as follows:}}
\begin{equation}
    \label{equ: v_hat 2-2}
\tilde{\bm V} = \begin{bmatrix}
 	e^{\iu\phi} & 0 \\
 	0 & 1
 \end{bmatrix}	
 \begin{bmatrix}
 	\cos(\psi) &\sin{\psi}\\
 	-\sin(\psi) & \cos(\psi)
 \end{bmatrix},
\end{equation}
\Copy{R3-3-2}{\addrev{where $\phi\!\in\![-\pi,\pi]$ and $\psi \!\in\! [0,\pi/2]$. 
Based on $\tilde{\bm V}= \bm R(\bm V)$ and the solved $\bm V$ from \eqref{equ: general equ}, we determine the close-form expressions of $\phi$ and $\psi$ by~\eqref{eq: 2-2 phi} and~\eqref{eq: 2-2 psi}, respectively, and thus derive the relationship between BFI and CSI.}}

\begin{figure}[t]
    \centering
    \includegraphics[width=.35\textwidth]{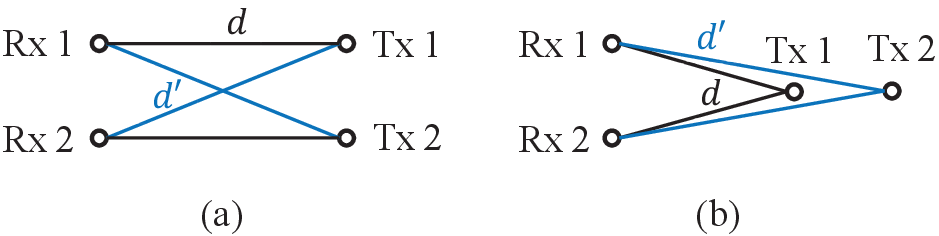}
    \vspace{-1em}
    \caption{Two special cases of $2\times 2$ MIMO systems, where $d$ and $d'$ represent distance between antennas.}
    \label{fig:2x2MIMO}
    \vspace{-1em}
\end{figure}

\Copy{R1-2nd-Q1}{To provide more insight, consider two special cases shown in \minirev{Fig.~\ref{fig:2x2MIMO}}.}
Based on~\eqref{eq: 2-2 phi} and~\eqref{eq: 2-2 psi}, it can be derived that for Case~(a), $\phi=\mod(2\pi(d-d')/\lambda,2\pi)$ and $\psi=\pi/4$; for Case~(b), $\phi=\mod(2\pi(d-d')/\lambda,2\pi)$ and $\psi = \tan^{-1}(\frac{2}{d'/d-d/d'})/2$.
Comparing the results for the two special cases, we observe that $\phi$ indicates the phase difference between a Tx to different Rxs, while $\psi$ implies the asymmetry of two Txs with respect to the Rxs.
The BFI elements generally contain distinct physical information about the wireless channels, confirming our analysis for the general case.
\begin{figure*}[b]
\centering
\vspace{-0em}
\rule{\linewidth}{0.4pt}
\beq
\label{eq: 2-2 phi}
\small
\phi = \chi = \tan^{-1}\left( \frac{\alpha_{11} \alpha_{12} \cdot \sin \left(2 \pi (t_{11} - t_{12}) \right) + \alpha_{21} \alpha_{22} \cdot \sin \left(2 \pi (t_{21} - t_{22}) \right)}{\alpha_{11} \alpha_{12} \cdot \cos \left(2 \pi (t_{11} - t_{12}) \right) + \alpha_{21} \alpha_{22} \cdot \cos \left(2 \pi (t_{21} - t_{22}) \right)} \right),
\eeq
\vspace{-.2em}
\beq
\label{eq: 2-2 psi}
\small
\psi = \mathrm{tan}^{-1}\left( \frac{\sqrt{4(\alpha_{11} \alpha_{12})^2 + 4(\alpha_{21} \alpha_{22})^2 + 8 \alpha_{11} \alpha_{12} \alpha_{21} \alpha_{22} \cdot \cos \left( 2 \pi (t_{11}-t_{12}-t_{21}+t_{22}) \right) }}{\alpha_{11}^{2} + \alpha_{21}^{2} - \alpha_{12}^{2} - \alpha_{22}^{2}} \right) / 2.
\eeq
\end{figure*}

With $\phi$ and $\psi$ as derived in~\eqref{eq: 2-2 phi} and~\eqref{eq: 2-2 psi}, it is evident that they encode both the phase and amplitude information of the CSI, despite being angular values.
A closer observation reveals that $\phi$ and $\psi$ only preserve the phase difference and amplitude ratio, rather than the original 
values recorded in CSI.
This observation can be extended to general cases,
aiding in comprehending the intrinsic information contained in BFI and its distinctions from CSI.

Furthermore, the complex nonlinear transformation from CSI to BFI results in variations of the sensitivity towards wireless channels and user positions. 
Specifically, as different elements in $\bm \theta$ are formed at diverse stages of the iterative transformation, i.e., $\bm g_{\sGR}(\cdot)$, differences in the amount of contained positional information are inevitable. 
To analyze the sensitivity of BFI towards user position and quantity the 
information of the BFI elements, in the following section, we derive an approximated CRB for the BFI.

\section{Efficient Feature Selection Algorithm for BFI-Based Sensing}
\label{sec: aprox BCRB}

In this section, we first propose an efficient approximation to calculate the CRB for BFI-based Wi-Fi sensing.
Based on this, we propose an efficient feature selection algorithm, facilitating the neural network pruning and parameter reduction for general BFI-based sensing.

\vspace{-1.5ex}
\subsection{Gaussian Kernel-Based CRB Approximation} \label{ssec: Gauss BCRB}

The CRB depicts the precision lower bound in terms of the expected mean-squared error.
Specifically, for a $D$-dim vector $\bm x = (x_1,...,x_{D})$ to be estimated, the covariance matrix for estimating $\bm x$ with $\hat{\bm x}(\bm \theta)$ can be calculated by
\beq
\label{equ: mse matrix}
\bm\varSigma(\bm x) = \mathbb E_{\bm \theta}\Big(
\big(\hat{\bm x}(\bm \theta)-\bm x)(\hat{\bm x}(\bm \theta) - \bm x)^{\top}
\Big),
\eeq
where $\bm \theta$ represents the vector of observation values dependent on $\bm x$, and $\hat{\bm x}(\bm \theta)$ is the estimation of $\bm x$ based on $\bm\theta$.

\Copy{R3-3-3}{\addrev{In the following, we derive an approximated CRB expression for BFI in Wi-Fi sensing based on the general CRB expressions in~\cite{hossain2010cramer}. Based on~\cite{hossain2010cramer}, the covariance matrix for $\bm x$ in~\eqref{equ: mse matrix} is lower bounded by $\bm \varSigma(\bm x) \succeq (\bm J(\bm x))^{-1}$.}} Here, $\bm J(\bm x)$ is the Fisher information matrix (FIM) defined by
\begin{align}
\label{equ: global FIM}
[\bm J(\bm x)]_{i,j} = &
- \mathbb E_{\bm\theta} \Big(
\frac{\partial^2 \ln p(\bm \theta|\bm x)}{\partial x_i\partial x_j}
\Big)  \\
= & -\int_{\bm \theta} \Big(
\frac{\partial \ln p(\bm \theta|\bm x)}{\partial x_i}
\Big) 
\Big(
\frac{\partial \ln p(\bm \theta|\bm x)}{\partial x_j}
\Big)
p(\bm \theta|\bm x) \dd \bm \theta, \nonumber
\end{align}
where 
$p(\bm \theta|\bm x)$ is the conditional probability density function~(PDF) of $\bm \theta$ given $\bm x$.

For a BFI-based Wi-Fi sensing system, the observed values are the BFI vector $\bm\theta$, and the parameter vector $\bm x$ indicate the positional information of the user, e.g., AoA, AoD, and/or the distance between the UD and AP.
To calculate the CRB, 
$p({\bm \theta}|\bm x)$ needs to be determined, which is highly challenging due to the complicated transformation from $\bm H$ to $\bm \theta$. 
Though existing work has derived the PDF of $\bm \theta$ under the strict assumption of a zero-mean Gaussian random channel, evaluating $p({\bm \theta}|\bm x)$ is still hard under general channel models~\cite{Godana2013TWC_Parametrization}.

To handle this issue, we propose to approximate $p({\bm \theta}|\bm x)$ with the multi-variant Gaussian kernel function.
This approximation is based on our simulation results in Sec.~\ref{ssec: Validation Gauss}, which reveal that, under
the condition of high SNR, each element of $\bm \theta$ approximately follows a Gaussian distribution. 
Therefore, as the SNR is generally high in wireless sensing scenarios of Wi-Fi networks, $p(\bm \theta|\bm x)$ can be approximated by

\beq
\label{equ: multivar gaussian}
p(\bm \theta|\bm x) \approx \frac{\exp \left(-\frac{1}{2}(\bm\sigma(\bm \theta, \bar{\bm \theta}))^{\top} {\bm C}^{-1}\bm\sigma(\bm \theta,\bar{\bm \theta})\right)}{\sqrt{(2 \pi)^{N_{\sG}}|\bm C|}},
\eeq
where $\bar{(\cdot)}$ denotes the expectation. $\bm C$ is the co-variance matrix of $\bm \theta$. These can be approximated numerically by Monte Carlo sampling.
Besides, in~\eqref{equ: multivar gaussian}, $\bm\sigma(\bm \theta, \bar{\bm \theta})$ represents difference between the elements of $\bm \theta$ and $\bar{\bm \theta}$ considering their value ranges and periodicity, whose $i$-th element~($i\in\{1,...,N_{\sBFI}\}$) is
\begin{equation}
\label{equ: period consider}
[\bm\sigma(\bm\theta,\bar{\bm\theta})]_i= 
\begin{cases}
(\theta_i - \bar{\theta}_i) - 2\pi \lfloor\frac{(\theta_i - \bar{\theta}_i)}{\pi}\rfloor ,~\text{if $\theta_i\in\bm\phi$},\\
(\theta_i - \bar{\theta}_i) - {\pi\over 2} \lfloor\frac{4(\theta_i - \bar{\theta}_i)}{\pi}\rfloor ,~\text{if $\theta_i\in\bm\psi$},\\
\end{cases}
\end{equation}
where $\lfloor \cdot \rfloor$ represents the floor function.

Equ.~\eqref{equ: period consider} implies that we only need to consider the shortest variation range of $\bm \theta$, thus avoiding abrupt gradient changes caused by crossing periods.
Besides, in~\eqref{equ: multivar gaussian}, expectation $\bar{\bm \theta}$ can be approximated as $\bar{\bm \theta} = \bm g_{\sGR}\circ\bm R\circ \bm F_{\srsvd}(\bar{\bm H})$ with $\bar{\bm H}$ being the expectation of $\bm H$.
Then, utilizing~\eqref{equ: multivar gaussian} and the CRB formula for multi-variant Gaussian observations~\cite[Eq.~(3.31)]{Kay1993Fundamentals}, the CRB of the $i$-th positional parameter, i.e., $x_i$~($i\in\{1,..., D\}$), can be approximated by
\beq
\label{equ: crb approx result}
[(\bm J(\bm x))^{-1}]_{i,i} = \left({\partial \bar{\bm \theta}/\partial x_i}\right)^{\top} \bm C^{-1} {\left(\partial \bar{\bm \theta}/ \partial x_i\right)}.
\eeq
\subsection{Feature Selection Algorithm for \rev{BFI-Based} Sensing} \label{ssec: feature selection algorithm}

\Copy{R2-1-a}{To reduce the computational burden on the edge device, we propose a CRB-based feature selection algorithm, where \addrev{$N_{\sSel}$, a user-defined value based on UD's resource limitation}, is selected from the $N_{\sBFI}$ BFI elements in each subcarrier as the features for BFI-based sensing.}
Intuitively, we can enumerate the summed CRB values for all possible combinations of $N_{\sSel}$ BFI elements w.r.t. the ROI and select the group of BFI elements with the lowest value.
Nevertheless, this method is of $\mathcal O(C_{N_{\sBFI}}^{N_{\sSel}})$ computational complexity, which can be large even for a relatively small $N_{\sBFI}$.
\Copy{R3-5-3}{Moreover, with a significant number of subcarriers, it would further incur prohibitive costs}. 
To handle this issue, we propose an efficient algorithm with $\mathcal O(N_{\sBFI})$ complexity.
Specifically, we first discretize the ROI into $R$ points and then calculate the CRB values of individual BFI elements at each point.
\Copy{R3-1-3}{Subsequently, we select each of the $N_{\sSel}$ BFI elements in a greedy manner, iteratively identifying the one that leads to the lowest CRB values for the greatest number of remaining points, and then removing these positions for the next iteration}.
The details are shown in Algorithm~\ref{alg: feature selection}. 

\begin{algorithm}[h]
    \caption{\small{Efficient CRB-Based BFI Feature Selection}}
    \KwIn{The $R$ points of the ROI ($\mathcal R$); the \addrev{user-defined} number of selected elements ($N_{\sSel}$).
    }
    
    \KwOut{
    Selected BFI elements $\left(\{\mathcal B_{1},...,\mathcal B_{N_{\ssc}}\} \right)$.
    }

    \vspace{.3em}
    
    \For{$k=1,...,N_{\ssc}$}
    {
    \vspace{.2em}
    Initialize $\mathcal B_{k}=\emptyset$, $\bm \eta=\bm 0$ ($\bm\eta\in \mathbb N^{R}$), $\chi_{\min}=+\infty$\;
    \vspace{.2em}
    \For{$\bm x\in \mathcal R$ and $j=1,..., N_{\sBFI}$}{
    Calculate the CRB of the $j$-th BFI element for $\bm x$ by
    $\chi_{j}\! =\! \sum_{i=1}^{D} \!|{\partial \bar{\theta}_{j}/\partial x_i}|_2^2/\delta_{k,j}$ based on~\eqref{equ: crb approx result}, \addrev{with $\delta_{k,j}$ being the variance of $\theta_j$} \;
    \vspace{.2em}
    \If{$\chi_{\min}>\chi_j$}{\addrev{Update the minimum CRB value and element index:} $\chi_{\min}=\chi_j$ and $\eta_{r}=j$\;}
    }
    \For{$j=1,...,N_{\sSel}$}{\addrev{Identify the element in the remaining set that has the lowest CRB values at most positions:}
    $\ell^* = \arg\max_{\ell\in\{1,...,N_{\sBFI}\}\setminus\mathcal B_k} \sum_r \mathbb I(\eta_{r}=\ell)$\;
    \vspace{.2em}
    Add $\ell^*$ to $\mathcal B_k$, i.e., $\mathcal B_k = \mathcal B_k \cup \{\ell^*\}$\;
    }
   }
  \label{alg: feature selection}
\end{algorithm}

\vspace{-2ex}

\section{Simulation Results} \label{sec:simu and evalu}

In this section, we evaluate the CRB of BFI-based Wi-Fi sensing as well as feature selection algorithm through extensive simulations.
After simulation setup, we first validate the PDF approximation of BFI elements in~\eqref{equ: multivar gaussian}.
Then, we compare the CRB of the BFI and CSI to show their respective sensing capabilities, and evaluate BFI-based sensing systems with different numbers of Tx-Rx pairs.
Finally, we verify the efficiency of Algorithm~\ref{alg: feature selection} through baseline comparisons.

\vspace{-2ex}
\subsection{Simulation Setup} \label{ssec:simu setup}
In the simulation, we consider a MIMO system with an AP and a UD. The UD is within a 2D circular area centered at the AP, and the distance range between the AP and the UD is $[5,10]$~\!m.
Initially, the UD's default position in terms of AoD, AoA, and distance is $(0^{\circ},0^{\circ},5~\!\text{m})$.
The center frequency of the system is  $5.825$~\!GHz, and, without loss of generality, we focus on sensing with the center subcarrier.
Besides, the Tx and Rx antennas of the AP and UD are linearly arranged, with a spacing interval equal to half the central wavelength.
The AP continuously transmits downlink packets to the UD. 
\Copy{R3-1-4}{The UD obtains the CSI from the LTF of the packets, calculates BFI, and sends the BFI to the AP}.
As for the channel model, the IEEE 802.11 TGn model of wireless channel is adopted.
Moreover, for the purpose of controlling variables, we assume the AP controls its transmit power to ensure a SNR of $20$~\!dB.

\vspace{-2ex}
\subsection{Validation of PDF Approximation for BFI} 
\label{ssec: Validation Gauss}

\begin{figure}[b]
    \centering
    \subfigure[]{
        \includegraphics[width=0.225\textwidth]{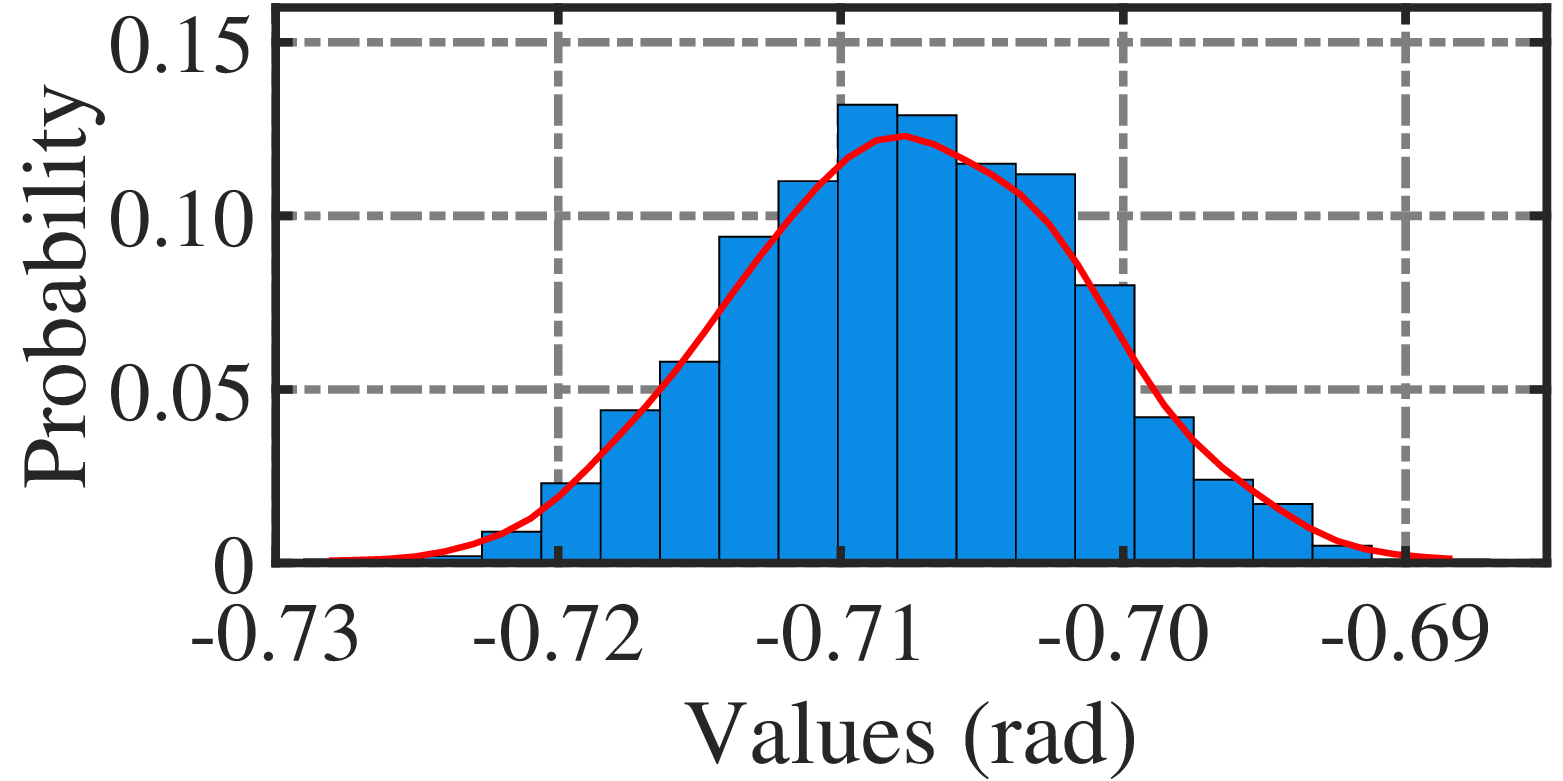}
        \label{sfig: guass_prove_phi11}
        \vspace{-2.5em}}
    \hspace{-2ex}
    \subfigure[]{
        \includegraphics[width=0.225\textwidth]{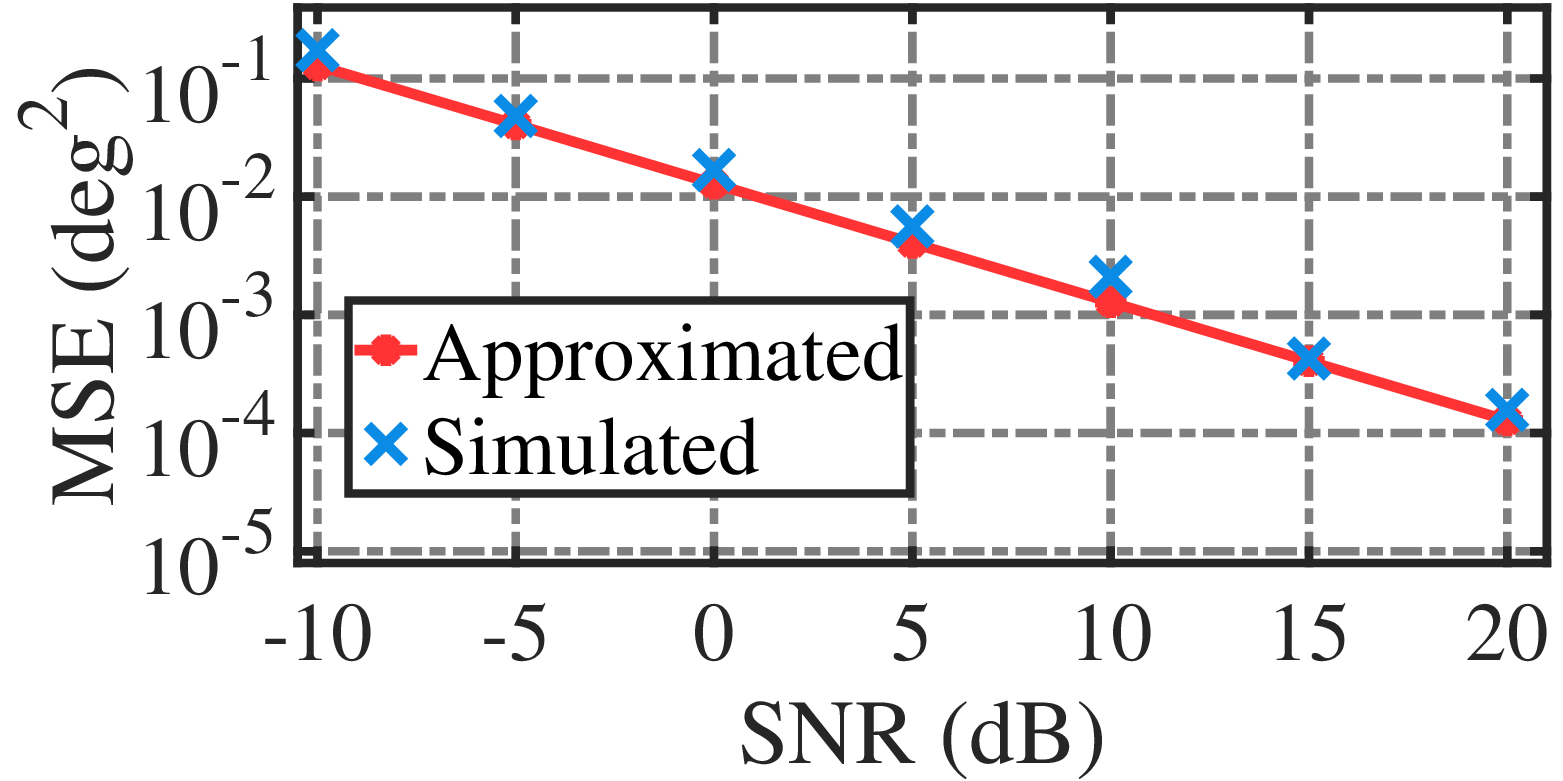}
        \label{sfig: prove_monte_carlo}
        \vspace{-2.5em}}
    \vspace{-1.5ex}
    \caption{(a) Comparison between the histogram of $\theta_1$ and Gaussian PDF; (b) Variances of the approximated CRB and the Monte Carlo simulation.}
    \vspace{-0.5ex}
    \label{fig: guass_prove}
\end{figure}

\Copy{R2-3-1}{We generate $10^4$ samples of CSI with random noise \addrev{in a system with 4 Tx and 4 Rx antennas} and compute their corresponding BFIs, where each BFI consists of \rev{$N_{\sBFI}=12$} elements.}
In Fig.~\ref{sfig: guass_prove_phi11}, we take $\theta_1$~($\phi_{11}$) as an example and illustrate its histogram, which fits well with the Gaussian distribution shown by the red curve.
Additionally, we calculate \emph{p-values} using the Kolmogorov–Smirnov test~\cite{smirnov1948table} to quantify the probability of observing the set of BFI elements under the hypothesis of a Gaussian distribution. The results show that all values exceed 0.8, validating our approximation in~\eqref{equ: multivar gaussian}.

To validate the calculated CRB, we employ~\eqref{equ: crb approx result} to compute the variance under various SNR conditions. We also conduct Monte Carlo simulations using the MUSIC algorithm as outlined in~\cite{Itahara2022Acc_Beamforming}. The results depicted in Fig.~\ref{sfig: prove_monte_carlo} reveal a close match in variances between the two approaches, affirming the viability of the proposed approximate CRB.

\vspace{-2ex}
\subsection{Comparison of \rev{BFI-Based} Sensing Capability} \label{ssec:comparison of sensing capability}

\begin{figure}[t]
    \centering
    \subfigure[]{
        \includegraphics[width=0.135\textwidth]{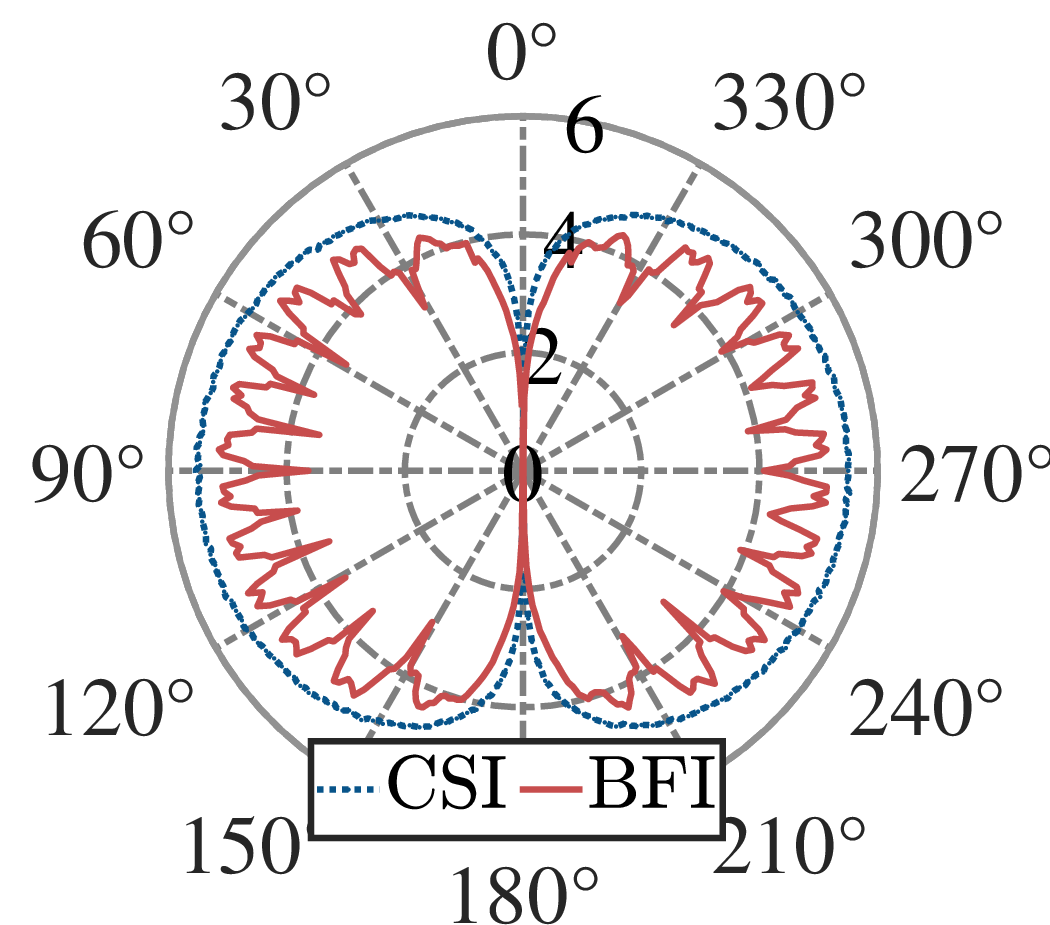}
        \label{sfig: AoD_los_process}
       \hfill}
    \hspace{-2.8ex}
    \subfigure[]{
        \includegraphics[width=0.135\textwidth]{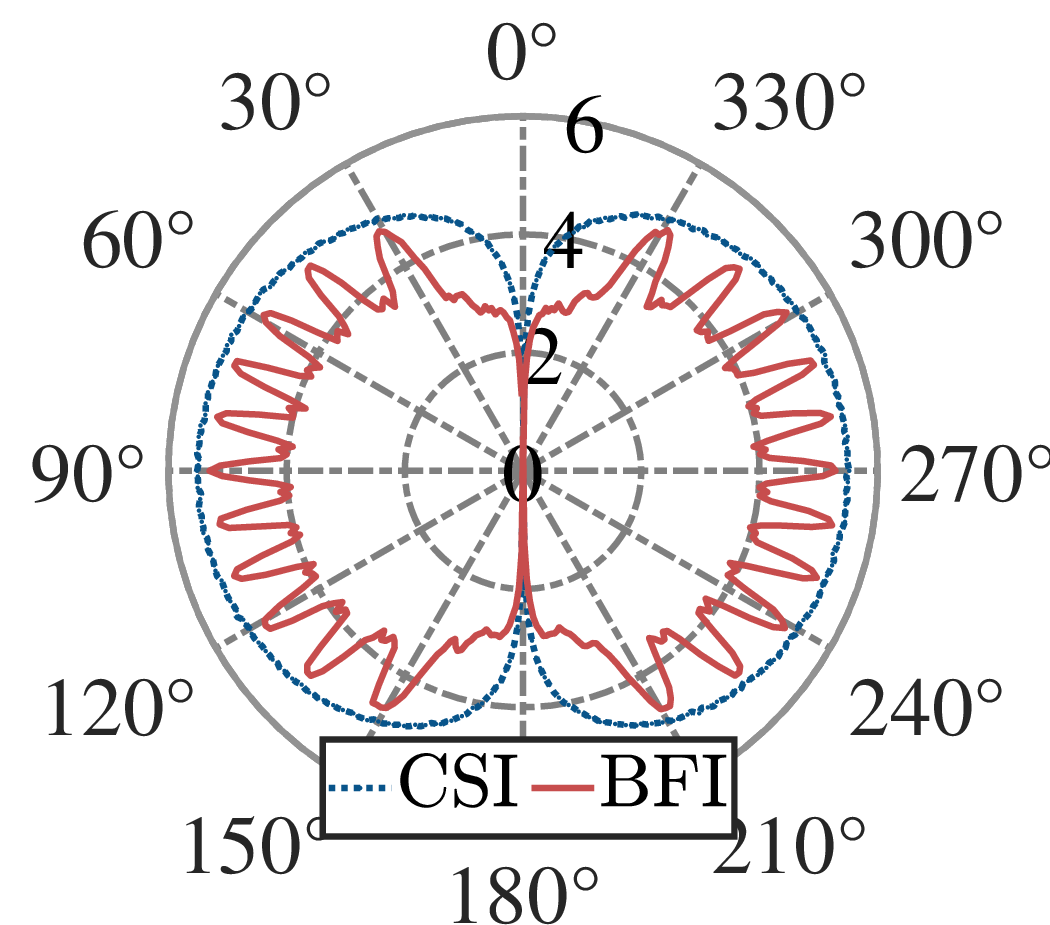}
        \label{sfig: AoA_los_process}
        \vspace{-2.2ex}}
    \hspace{-2.5ex}
    \subfigure[]{
        \includegraphics[width=0.170\textwidth]{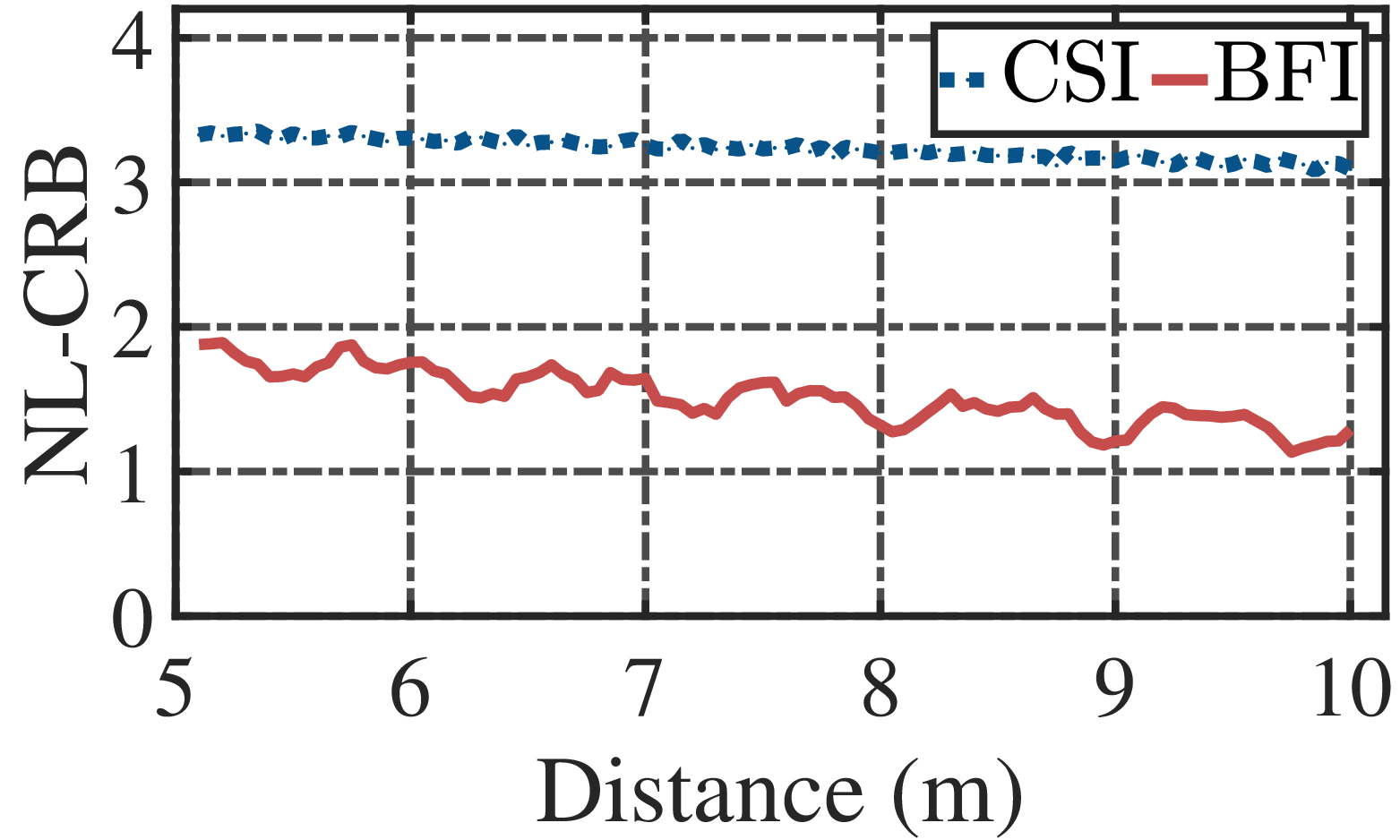}
        \label{sfig: ToF_los_process}
        \hfill}
    \vspace{-1.5ex}
    \caption{Comparison of sensing capability between CSI and BFI in terms of the NL-CRB for (a) AoD, (b) AoA, and (c) distance.}
    \label{fig: CSI_BFI_compare}
\end{figure}

\begin{figure}[t]
    \centering
    \subfigure[]{
        \includegraphics[width=0.135\textwidth]{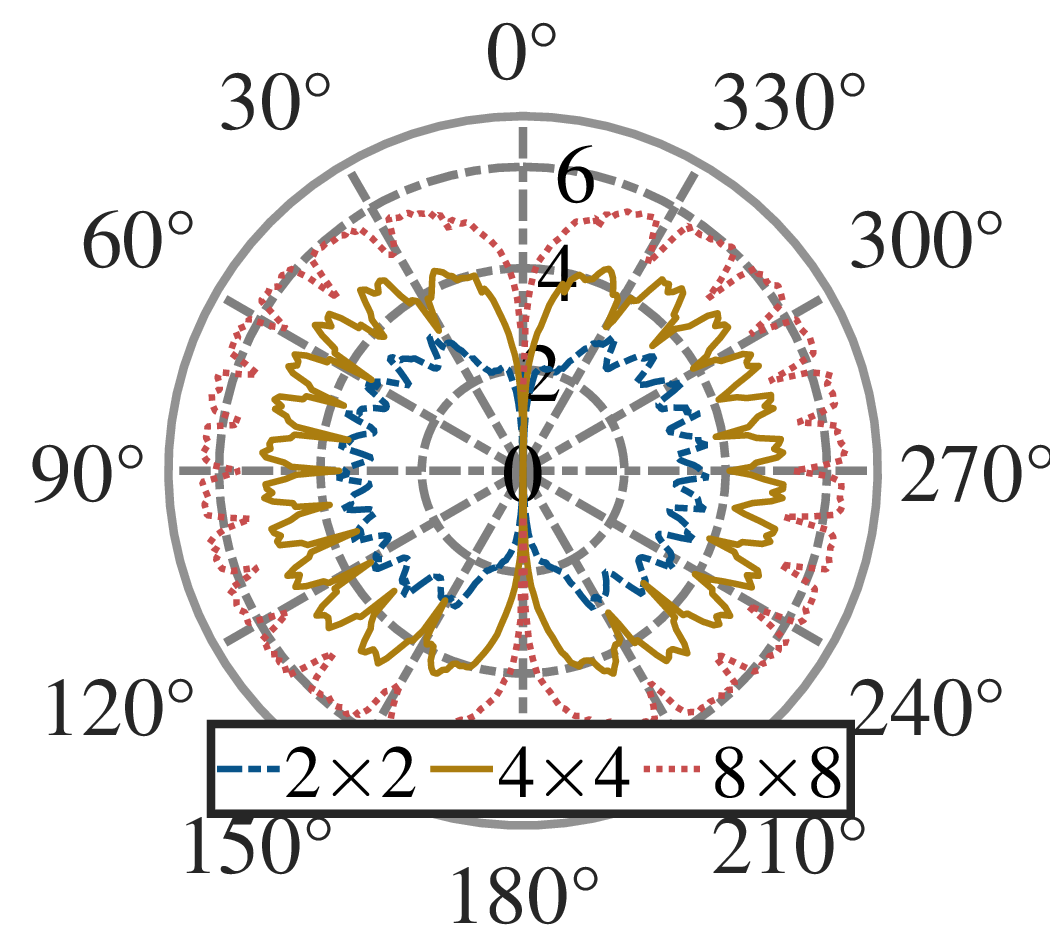}
        \label{sfig: AoD_los_MIMO}
        \hfill}
    \hspace{-2.8ex}
    \subfigure[]{
        \includegraphics[width=0.135\textwidth]{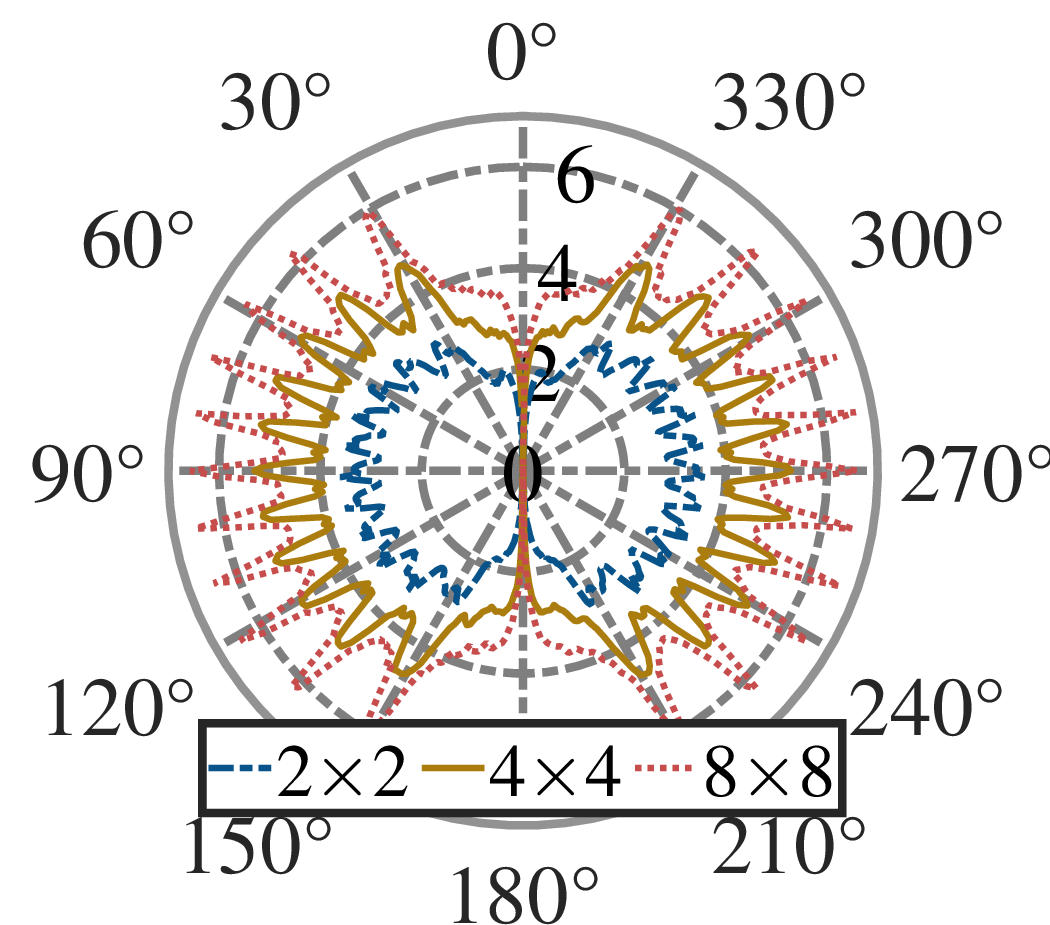}
        \label{sfig: AoA_los_MIMO}
        \vspace{-2.2ex}}
    \hspace{-2.5ex}
    \subfigure[]{
        \includegraphics[width=0.170\textwidth]{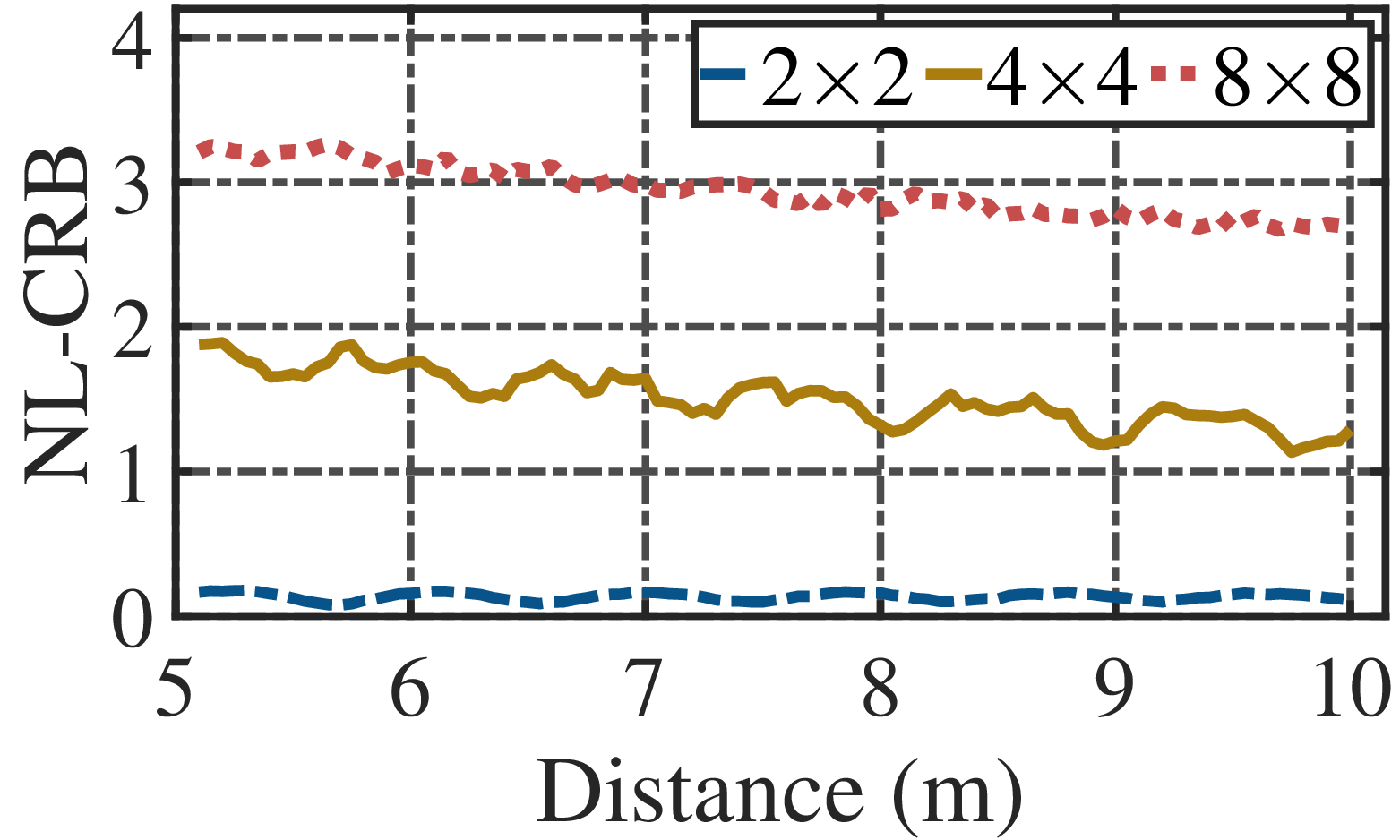}
        \label{sfig: ToF_los_MIMO}
    \hfill}
    \vspace{-1.5ex}
    \caption{Comparison of sensing capability among BFI in different MIMO systems in terms of the NL-CRB for (a) AoD, (b) AoA, and (c) distance.}
    \vspace{-2ex}
    \label{fig: CSI_MIMO_compare}
\end{figure}

To visualize the CRB clearly and intuitively, we illustrate the negative logarithm of the CRB~(NL-CRB), which is positively related to the sensing capability and facilitates the observation of CRB values with different orders of magnitude.
In Figs.~\ref{sfig: AoD_los_process},~\ref{sfig: AoA_los_process}, and~\ref{sfig: ToF_los_process}, we compare the sensing capabilities of BFI and CSI in terms of AoA, AoD, and distance, respectively.
It can be observed that, compared to CSI, BFI contains comparable positional information and can support sensing precision at a similar level.
\Copy{R3-1-5}{Nevertheless, it is also evident that BFI-based sensing exhibits decreased accuracy and stability compared to CSI-based sensing, which is due to its composition of only part of CSI decomposition}.
Besides, in Figs.~\ref{sfig: AoD_los_MIMO}, \ref{sfig: AoA_los_MIMO}, and \ref{sfig: ToF_los_MIMO}, we compare the sensing capabilities of BFI in MIMO systems with different numbers of antennas.
It is shown that as the numbers of Tx and Rx antennas grow, the sensing capability increases rapidly, and the variance of sensing precision is also magnified, which is probably due to the enhanced directionality of larger antenna arrays.

\vspace{-1.0ex}

\subsection{Evaluation of \rev{CRB-Based} Feature Selection}  \label{ssec: evalu feature selection}
\begin{figure}[b]
    \vspace{-2ex}
    \centering
    \includegraphics[width=0.456\textwidth]{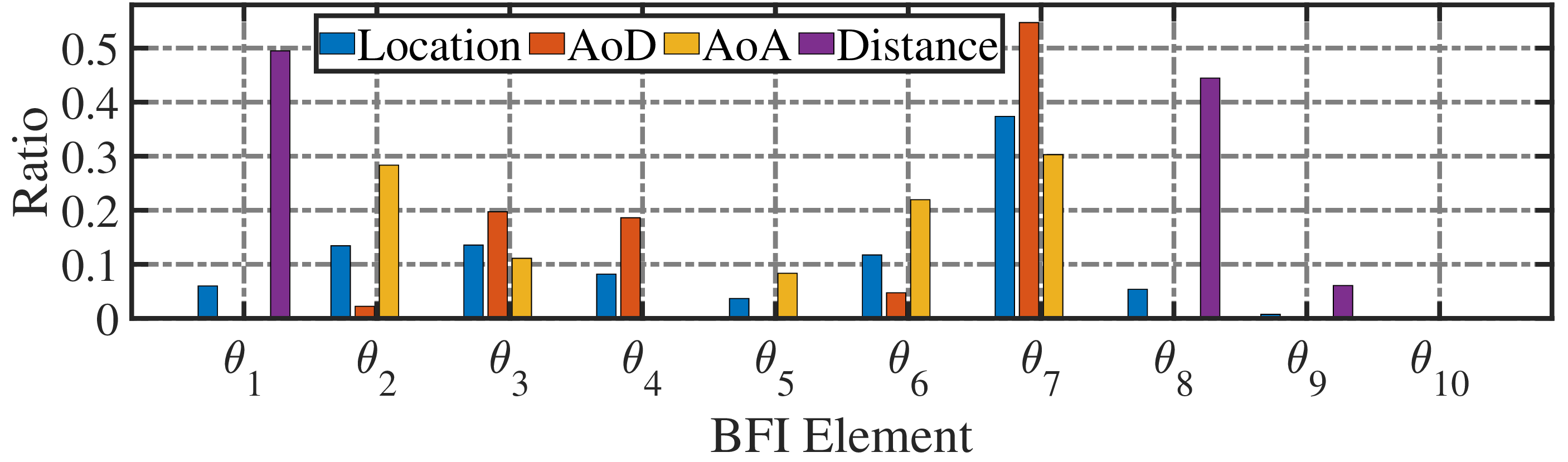}
    \vspace{-2ex}
    \caption{\Copy{R3-3-4}{\addrev{Ratio of the positions where each $\theta_i$ has the lowest CRB among all BFI elements, given the cases for sensing location, AoD, AoA, and distance.}}}
    \label{fig: rate_lowest_all}
\end{figure}

\begin{figure}[t]
    \centering
    \subfigure[]{
        \includegraphics[width=0.225\textwidth]{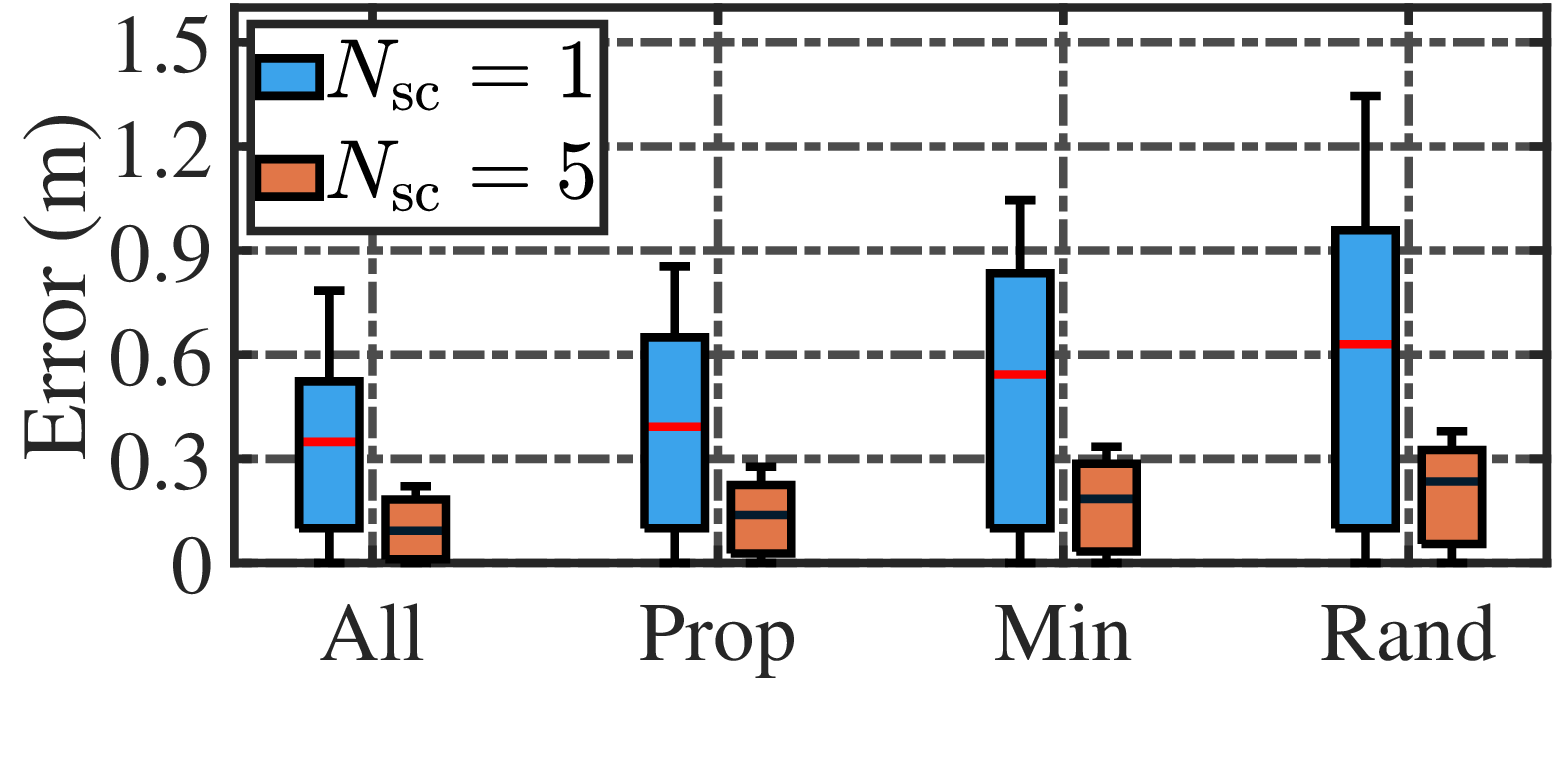}
        \label{sfig: box_method}
        \vspace{-3em}}
    \subfigure[]{
        \includegraphics[width=0.225\textwidth]{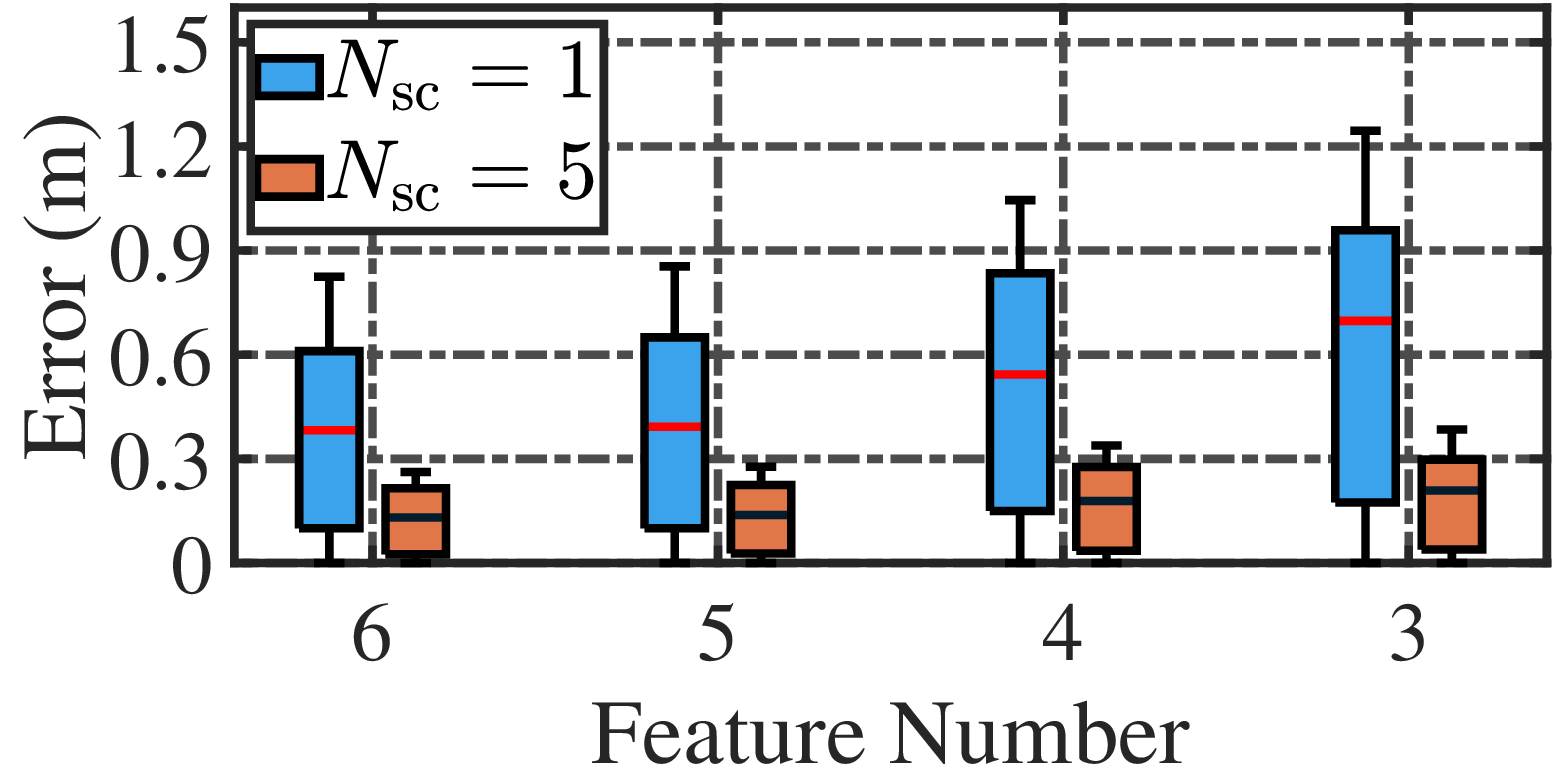}
        \label{sfig: box_number}
        \vspace{-3em}}
    \vspace{-2.0ex}
    \caption{\addrev{Comparison of the distributions of positioning errors for (a) different feature selection algorithms given $N_{\sSel}=5$ and (b) different $N_{\sSel}$ in the proposed Algorithm~\ref{alg: feature selection}.}}
    \vspace{-2ex}
    \label{fig: comparison_feature_selection}
\end{figure}

\Copy{R2-3-2}{We consider an MIMO system comprised of an AP with $4$ Tx antennas and a UD with $2$ Rx antennas \addrev{(resulting in 10 BFI elements)}, \addrev{which is prevalent  in daily scenarios~\cite{hu2023muse}.}}
\Copy{R2-4}{We assume a three-layered multilayer perceptrons~(MLPs) \addrev{with $N_{\sSel}\times N_{\ssc}$ input neurons} is employed to map BFI to UD's position.} In this case, we analyze $R=10^3$ positions.
\Copy{R2-2nd-Q1}{\minirev{Our evaluation in Fig.~\ref{fig: rate_lowest_all} clearly shows that the performance of each BFI element varies with different parameters to be sensed, i.e., the location, AoD, AoA, or distance.}}
Here, each bar represents the ratio between the number of discrete points where the BFI element has the lowest CRB value to the total number of points.
It can be observed in Fig.~\ref{fig: rate_lowest_all} that different sets of BFI elements need to be selected when sensing the UD's 2D location, AoA, AoD, and distance, respectively. 
\Copy{R2-2-a}{\addrev{We note that the zero ratio of $\theta_{10}$ indicates that $\theta_{10}$ does not possess the lowest CRB at all the positions, rather than it contains no positional information.}}

To evaluate the proposed Algorithm~\ref{alg: feature selection}~(\textbf{Prop}), we compare it with three baseline algorithms, including: 1) selecting all the features~(\textbf{All}), 2) selecting the $N_{\sSel}$ BFI elements with the minimal average CRB values~(\textbf{Min}), and 3) selecting the $N_{\sSel}$ BFI elements randomly~(\textbf{Rand}).
In Fig.~\ref{sfig: box_method}, it can be observed that using half of the BFI elements selected by our proposed algorithm~(i.e., $N_{\sSel}=5$), the trained MLP achieves positioning errors close to those for the MLP using all BFI elements.
The proposed algorithm selects the BFI elements that can support positioning with lower errors (outperforms the second best by \addrev{over $25$\%} in terms of median values) and are more efficient compared to the other baselines.
\Copy{R2-Q1-a-2}{\addrev{Furthermore, users can also determine $N_{\sSel}$ after trials of its different values with the help of Algorithm~\ref{alg: feature selection}.}}
\Copy{R3-5-5}{We compare the positioning errors of the MLPs using varying numbers of selected BFI elements across different subcarrier numbers $N_{\ssc}$, as shown in Fig.~\ref{sfig: box_number}. 
With our proposed algorithm, the MLP can achieve a median positioning error comparable to other baselines using only $4$ BFI elements as features, saving $20$\% of parameters}.

\section{Conclusions \addrev{and Outlook}} \label{sec:Conclusion}
In this letter, we have established the mathematical model of BFI and derived the closed-form BFI expression for $2\times 2$ MIMO systems.
We then analyzed the CRB of BFI using a Gaussian-kernel-based approximation and proposed an efficient BFI feature selection algorithm based on it.
Simulation results have shown that the BFI exhibits comparable sensing capability to CSI, and that the proposed algorithm can significantly reduce the number of required BFI features for UD positioning, outperforming baseline methods by over $20$\%.
\Copy{R1-Q1}{\addrev{In future studies, we will validate the \huv{proposed algorithm} in real-world environments and \huv{various} downstream applications using \huv{BFI collected by} commercial Wi-Fi NICs.}}

\bibliographystyle{IEEEtran}

\begin{thebibliography}{10}
\providecommand{\url}[1]{#1}
\csname url@samestyle\endcsname
\providecommand{\newblock}{\relax}
\providecommand{\bibinfo}[2]{#2}
\providecommand{\BIBentrySTDinterwordspacing}{\spaceskip=0pt\relax}
\providecommand{\BIBentryALTinterwordstretchfactor}{4}
\providecommand{\BIBentryALTinterwordspacing}{\spaceskip=\fontdimen2\font plus
\BIBentryALTinterwordstretchfactor\fontdimen3\font minus \fontdimen4\font\relax}
\providecommand{\BIBforeignlanguage}[2]{{%
\expandafter\ifx\csname l@#1\endcsname\relax
\typeout{** WARNING: IEEEtran.bst: No hyphenation pattern has been}%
\typeout{** loaded for the language `#1'. Using the pattern for}%
\typeout{** the default language instead.}%
\else
\language=\csname l@#1\endcsname
\fi
#2}}
\providecommand{\BIBdecl}{\relax}
\BIBdecl

\bibitem{hu2023muse}
J.~Hu, T.~Zheng, Z.~Chen, H.~Wang, and J.~Luo, ``{MUSE-Fi: Contactless MUti-person SEnsing Exploiting Near-field Wi-Fi Channel Variation},'' in \emph{Proc. ACM Mobicom}, 2023, pp. 1--15.

\bibitem{hu2023password}
J.~Hu, H.~Wang, T.~Zheng, J.~Hu, Z.~Chen, H.~Jiang, and J.~Luo, ``{Password-Stealing without Hacking: Wi-Fi Enabled Practical Keystroke Eavesdropping},'' in \emph{Proc. of 30th ACM CCS}, 2023, pp. 239--252.

\bibitem{Itahara2022Acc_Beamforming}
S.~Itahara, S.~Kondo, K.~Yamashita, T.~Nishio, K.~Yamamoto, and Y.~Koda, ``Beamforming feedback-based model-driven angle of departure estimation toward legacy support in wifi sensing: An experimental study,'' \emph{IEEE Access}, vol.~10, pp. 59\,737--59\,747, Jun. 2022.

\bibitem{Wu2023SIGCOMM_Enabling}
C.~Wu, X.~Huang, J.~Huang, and G.~Xing, ``Enabling ubiquitous wifi sensing with beamforming reports,'' in \emph{Proc. ACM SIGCOMM}, New York, NY, USA, Sep. 2023.

\bibitem{Jiang2022WCL_On}
Y.~Jiang, X.~Zhu, R.~Du, Y.~Lv, T.~X. Han, D.~X. Yang, Y.~Zhang, Y.~Li, and Y.~Gong, ``{On the Design of Beamforming Feedback for Wi-Fi Sensing},'' \emph{IEEE Wireless Commun. Lett.}, vol.~11, no.~10, pp. 2036--2040, Oct. 2022.

\bibitem{IEEE80211acD31}
\BIBentryALTinterwordspacing
{IEEE Standards Association}. (2012, August) Local and metropolitan area networks—specific requirements; part 11: Wireless lan medium access control (mac) and physical layer (phy) specifications; amendment 4: Enhancements for very high throughput for operation in bands below 6 ghz. IEEE P802.11ac/D3.1. [Online]. Available: \url{http://www.ieee802.org/11/private/Draft_Standards/11ac/DraftP802.11ac_D3.1.pdf}
\BIBentrySTDinterwordspacing

\bibitem{Kim2006VTC_Efficient}
J.~Kim and C.~Aldana, ``{Efficient feedback of the channel information for closed-loop beamforming in WLAN},'' in \emph{Proc. IEEE VTC}, Melbourne, Australia, May 2006.

\bibitem{hossain2010cramer}
A.~M. Hossain and W.-S. Soh, ``{Cramer-Rao Bound Analysis of Localization Using Signal Strength Difference as Location Fingerprint},'' in \emph{Proc. IEEE INFOCOM}, 2010, pp. 1--9.

\bibitem{Godana2013TWC_Parametrization}
B.~E. Godana and T.~Ekman, ``{Parametrization based limited feedback design for correlated MIMO channels using new statistical models},'' \emph{IEEE Trans. Wireless Commun.}, vol.~12, no.~10, Oct. 2013.

\bibitem{Kay1993Fundamentals}
S.~M. Kay, \emph{Fundamentals of Statistical Signal Processing: Estimation Theory}.\hskip 1em plus 0.5em minus 0.4em\relax Prentice-Hall, Inc., 1993.

\bibitem{smirnov1948table}
N.~Smirnov, ``Table for estimating the goodness of fit of empirical distributions,'' \emph{Ann. Math. Statist.}, vol.~19, no.~2, pp. 279--281, 1948.

\end{thebibliography}


\end{document}